\documentclass[10pt,twocolumn,letterpaper]{article}

\usepackage[pagenumbers]{cvpr} 

\definecolor{cvprblue}{rgb}{0.21,0.49,0.74}
\usepackage{graphicx}
\usepackage{amssymb}
\usepackage{xcolor}
\usepackage[misc]{ifsym}
\usepackage[pagebackref,breaklinks,colorlinks,allcolors=cvprblue,urlcolor=magenta]{hyperref}
\newcommand\blfootnote[1]{%
    \begingroup
    \renewcommand\thefootnote{}\footnote{#1}%
    \addtocounter{footnote}{-1}%
    \endgroup
}
\usepackage[hang]{footmisc}

\title{EchoMimicV2: Towards Striking, Simplified, and Semi-Body Human Animation}
\author{Rang Meng\textsuperscript{\rm \dag}\textsuperscript{\rm \ddag} \quad
Xingyu Zhang \quad
Yuming Li\textsuperscript{\rm \ddag} \quad
Chenguang Ma\textsuperscript{\rm \ddag}
\vspace{2mm}
\\
\fontsize{11pt}{13pt}\selectfont{Terminal Technology Department, Alipay, Ant Group}
\\
{\tt\small
\{mengrang.mr, luoque.lym, chenguang.mcg\}@antgroup.com}
\vspace{1.5mm}
\\
\fontsize{10pt}{13pt}\selectfont{code: \href{https://github.com/antgroup/echomimic_v2}{\textcolor{magenta}{https://github.com/antgroup/echomimic\_v2}}}
}

\begin{document}
\maketitle
{
\blfootnote{
    {\rm \dag} Core contributor: coding, data curation, training, writing \& revision \\
    {\rm \ddag} Corresponding author}
}

\begin{abstract}
Recent work on human animation usually involves audio, pose, or movement maps conditions, thereby achieving vivid animation quality. However, these methods often face practical challenges due to extra control conditions, cumbersome condition injection modules, or limitations to head region driving. Hence, we ask if it is possible to achieve striking half-body human animation while simplifying unnecessary conditions. To this end, we propose a half-body human animation method, dubbed \textbf{EchoMimicV2}, that leverages a novel \textbf{Audio-Pose Dynamic Harmonization} strategy, including \textbf{Pose Sampling} and \textbf{Audio Diffusion}, to enhance half-body details, facial and gestural expressiveness, and meanwhile reduce conditions redundancy. To compensate for the scarcity of half-body data, we utilize \textbf{Head Partial Attention} to seamlessly accommodate headshot data into our training framework, which can be omitted during inference, providing a free lunch for animation. Furthermore, we design the \textbf{Phase-specific Denoising Loss} to guide motion, detail, and low-level quality for animation in specific phases, respectively.
Besides, we also present a novel benchmark for evaluating the effectiveness of half-body human animation. Extensive experiments and analyses demonstrate that EchoMimicV2 surpasses existing methods in both quantitative and qualitative evaluations. 
\end{abstract}    
\vspace{-1.5em}
\section{Introduction}
\label{sec:intro}

As deep learning develops, diffusion-based video generation has seen significant advancements\cite{dhariwal2021diffusion,ho2020denoising,rombach2022high,dhariwal2021diffusion,ho2020denoising,rombach2022high,guo2023animatediff,chen2023videocrafter1,blattmann2023stable,esser2023structure,yang2023rerender,brooks2024video,ignatov2018pirm,meng2022attention,meng2022slimmable,meng2020neural}, prompting extensive research in human animation.

Human animation, as a subset of video generation domain, aims at synthesizing natural and realistic human-centric video sequences from a reference character images. 

Current research on human animation commonly employs pretrained diffusion models as backbone, and involves corresponding condition injection modules to introduce control conditions\cite{chen2024echomimic,hu2024animate,zhang2024mimicmotion,wang2024v,wei2024aniportrait,xu2024hallo,zhang2023sadtalker,lin2024cyberhost}, so that lifelike animation can be well generated. Unfortunately, there
is still gaps between academic study and industrial needs: 1) Head region limitation; 2) Condition-injection complexity.
1) \textbf{Head Region Limitation}. On the one hand, prior human animation works\cite{chen2024echomimic,wei2024aniportrait,xu2024hallo,zhang2023sadtalker} mainly focus on generating head-region videos, neglecting the synchronization of the audio and the shoulders-below body. The most recent work\cite{lin2024cyberhost} has improved half-body animation with auxiliary conditions and injection modules beyond audio-driven module. 
2) \textbf{Condition-Injection Complexity}. On the other hand, the commonly-used control conditions (\eg text, audio, pose, optical flow, movement maps) can provide a solid foundation for lifelike animation. 

Particularly, current research efforts concentrate on aggregating supplementary conditions, which result in unstable training due to multi-condition incoordination, and elevated inference latency stemming from intricate condition-injection modules.

For the first challenge, a straightforward baseline exists to accumulate conditions related to shoulder-below body, such as half-body key points maps. However, we discover that this approach remains infeasible because of increased complexity of conditions (for the second challenge).
 
In this paper, to remedy the aforementioned issues, we introduce an novel end-to-end method-EchoMimicV2, building upon the portrait animation method EchoMimic \cite{chen2024echomimic}. Our proposed EchoMimicV2 strives for striking quality of half-body animation yet with simplified conditions.

To this end, EchoMimicV2 exploits the Audio-Pose Dynamic Harmonization (APDH) training strategy to modulate both audio and pose conditions, and meanwhile reduce the redundancy of the pose condition. Additionally, it utilizes a stable training objective function, termed Phase-specific Loss (PhD Loss), to enhance motion, details, and low-level quality, replacing the guidance of redundant conditions.

Specifically, APDH is inspired by the waltz dance step, where audio and pose condition perform as synchronized dance partners. As the pose gracefully steps back, the audio seamlessly advances, perfectly filling in the space to create a harmonious melody. As a result, the control scope of the audio condition is extended from the mouth to the entire body via Audio Diffusion, and meanwhile the pose condition is confined from the entire body to the hands via Pose Sampling. 
Given that the primary regions responsible for audio expression located in the mouth lips, we initiate our audio condition diffusion from the mouth part. Additionally, due to the complementarity of gestural and verbal communication, we retain hand pose condition for the gesture animation so that the Head Region Limitation challenge is overcome, extending to half-body region animation.

Throughout this process, we find a free lunch for data augmentation. When audio condition only controls the head region via Head Partial Attention, we can seamlessly incorporate padded headshots data to enhance facial expressions without requiring additional plugins like\cite{lin2024cyberhost}. We also list the advantages of our proposed EchoMimicV2 in Table \ref{table_00}.

Moreover, we propose a stable training objective function, Phase-specific Loss (PhD Loss), with two goals: 1) to enhance the motion representation with incomplete pose; 2) to improve details and low-level visual quality not governed by audio. While it is intuitive to employ a multi-loss mechanism that integrates pose, detail, and low-level visual objective functions concurrently, such an approach typically requires extra models, including Pose Encoders and VAE Decoders. Given that the ReferenceNet-based backbone already demands significant computational resources\cite{hu2024animate}, implementing a multi-losses training becomes impractical.
Through experimental analysis, we segments the denoising process into three distinct phases, each with its primary focus:  1) Pose-dominant phase, where motion poses and human contours are initially learned; 2) Detail-dominant phase, where character-specific details are refined; and 3) Quality-dominant phase, where the model enhances the color and other low-level visual qualities. Consequently, the proposed PhD Loss is tailored to optimize the model for each specific denoising phase, that is, Pose-dominant Loss for the early phase, Detail-dominant Loss for the middle phase, and Low-level Loss for the final phase, ensuring a more efficient and stable training process.

Additionally, to facilitate the community in quantitative evaluation of half-body human animation, we have curated a test benchmark, named EMTD, comprising half-body human videos sourced from the Internet. We conducted extensive qualitative and quantitative experiments and analyses, demonstrating that our method achieves state-of-the-art results.
\begin{table}[t]
\begin{center}
\resizebox{.45\textwidth}{!}{
\begin{tabular}{ll}
\hline\hline
\multicolumn{1}{c|}{EchoMimicV2}                              & \multicolumn{1}{c}{CyberHost\cite{lin2024cyberhost}}                   \\ \hline
\multicolumn{2}{c}{Audio+RefImage}                                       \\ \hline
\multicolumn{1}{c|}{+ Hand Pose Sequence} & + Body Movement Sequence     \\
\multicolumn{1}{c|}{-}                    & + Face Crop Injection Module  \\
\multicolumn{1}{c|}{-}                    & + Hand Crop Injection Module \\
\multicolumn{1}{c|}{-}                    & + Full-body Pose Guidence    \\ 
\hline\hline
\end{tabular}
}
\end{center}
\vspace{-1.5em}
\caption{The simplification of our proposed EchoMimicV2.}
\vspace{-1.8em}
\label{table_00}
\end{table}
In summary, our contributions are as follows:
\begin{itemize}[leftmargin=12pt, topsep=2pt, itemsep=0pt]
\item We propose EchoMimicV2, an end-to-end audio-driven framework to generate striking half-body human animation yet driven by simplified conditions;
\item We propose APDH strategy to meticulously modulate audio and pose condition, and reduce pose condition redundancy;
\item We propose HPA, a seamlessly data augmentation to enhance the facial expressions in half-body animations, no need for additional modules;
\item We propose PhD Loss, a novel objective function to enhance the motion representation, appearance details and low-level visual quality, alternating the guidance of complete pose condition; 
\item We provide a novel evaluation benchmark for half-body human animation.
\item Extensive experiments and analyses are conducted to verify the effectiveness of our proposed framework, which surpasses other state-of-the-arts methods.
\end{itemize}

\begin{figure*}[t!]
\begin{center}
\includegraphics[width=.9\textwidth]{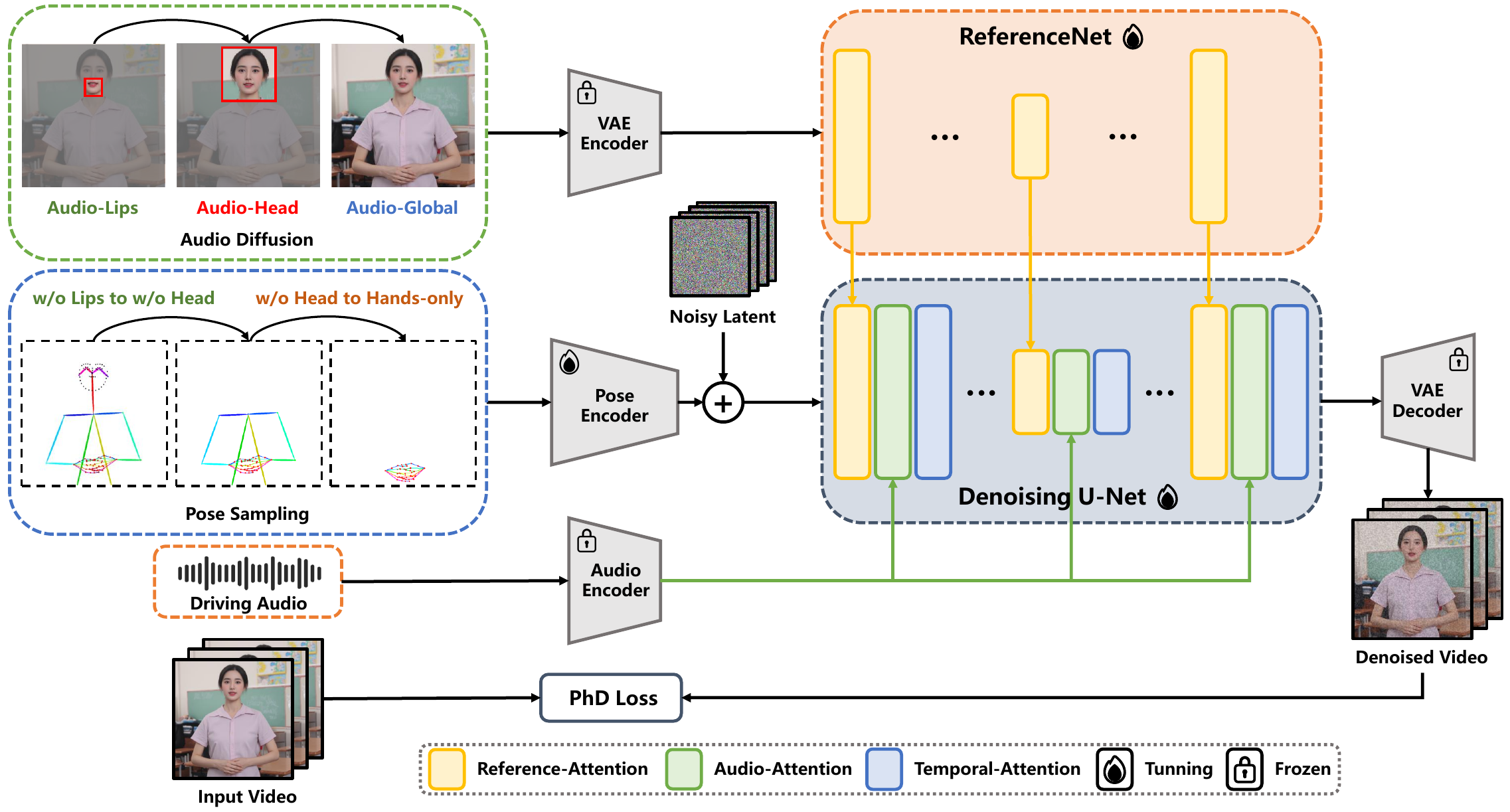}
\end{center}
\vspace{-1em}
\caption{The overall pipeline of our proposed EchoMimicV2.
}
\vspace{-1em}
\label{fig_02}
\end{figure*}
\section{Related Work}
\subsection{Pose Driven Human Animation}

Many human animation methods focus on pose-driven settings, using pose sequences from driving video. Research in pose-driven human video generation typically follows a standard pipeline, employing various pose types, such as skeleton pose, dense pose, depth, mesh, and optical flow—as guiding modalities, alongside text and speech inputs. Conditional generation models have advanced with stable diffusion (SD) or Stable Video Diffusion (SVD) as their backbone.
Methods like MagicPose\cite{chang2023magicpose} integrate pose features into diffusion models via ControlNet\cite{zhang2023adding}, while methods like AnimateAnyone\cite{hu2024animate}, MimicMotion\cite{zhang2024mimicmotion}, MotionFollower\cite{tu2024motionfollower}, and UniAnimate\cite{wang2024unianimate} extract skeleton poses using DWPose\cite{yang2023effective} or OpenPose\cite{qiao2017real}. These methods use lightweight neural networks with a few convolutional layers as pose guides to align skeleton poses with latent space noise during denoising.
In contrast, DreamPose\cite{karras2023dreampose} and MagicAnimate\cite{xu2024magicanimate} use DensePose\cite{guler2018densepose} to extract dense pose information, which is concatenated with noise and input into the denoising UNet by ControlNet\cite{zhang2023adding}. Human4DiT\cite{shao2024human4dit}, inspired by Sora, extracts 3D mesh maps using SMPL\cite{bogo2016keep} and adopts the Diffusion Transformer for video generation.
\subsection{Audio Driven Human Animation}

The goal of audio-driven human animation is to generate gestures from speech audio, ensuring motion aligns with audio in semantics, emotion, and rhythm. Existing works often focus on talking heads. EMO\cite{tian2024emo} introduces a Frame Encoding module for consistency. AniPortrait\cite{wei2024aniportrait} advances 3D facial structures mapped to 2D poses for coherent sequences. V-Express synchronizes audio with lip movements and expressions, refining emotional nuances. Hallo\cite{xu2024hallo} uses diffusion models for enhanced control over expressions and poses. Vlogger\cite{zhuang2024vlogger} generates talking videos from a single image via a diffusion model for high-quality, controllable output. MegActor-$\Sigma$\cite{yang2024megactor} integrates audio and visual signals into portrait animation with a conditional diffusion transformer. TANGO\cite{liu2024tango} generates co-speech body-gesture videos, improving alignment and reducing artifacts. Additionally, EchoMimic\cite{chen2024echomimic} is capable of generating portrait videos not only by audios and facial poses individually, but also by a combination of both audios and selected facial poses. CyberHost\cite{lin2024cyberhost} supports combined control signals from multiple modalities, including audio,full-body keypoints map, 2D hand pose sequence, and body movement maps.

\section{Method}
\subsection{Preliminaries}
\textbf{Latent Diffusion Model}. Our approach builds upon the Latent Diffusion Model (LDM)\cite{rombach2022high}, which employs a Variational Autoencoder (VAE) Encoder\cite{kingma2013auto} $E$ to map an image $I$ from the pixel sapce to a more compact latent space, represented as $z_0 = E(I)$. During training, Gaussian noise is progressively added to $z_0$ across various timesteps $t \in [1, \ldots, T]$, ensuring that the final latent representation $z_T$ follows a standard normal distribution $\mathcal{N}(0, 1)$. The primary training objective of LDM is to estimate the noise introduced at each timestep $t$, 
\begin{equation}
\begin{aligned}
L_{latent} = \mathbb{E}_{z_t, t, c, \epsilon \sim \mathcal{N}(0,1)}[\|\epsilon-\epsilon_\theta (z_t, t, c) \|_2^2]
\end{aligned}
\end{equation}
where $ \epsilon_\theta $ represents the trainable Denoising U-Net, and $c$ denotes the conditions like audio or text. During the inference phase, the pretrained model is used to iteratively denoise a latent vector sampled from a Gaussian distribution. The denoised latent is subsequently decoded back to an image using the VAE Decoder $D$.\\
\textbf{ReferenceNet-based Backbone}. As EchoMimic\cite{chen2024echomimic} and other  prevalent portrait animation work\cite{wei2024aniportrait,xu2024hallo,zhang2023sadtalker}, we exploit the ReferenceNet-based diffusion architecture as our backbone. In the ReferenceNet-based backbone, we utilize a duplicate of the pretrained 2D U-Net as the ReferenceNet to extract reference features from the reference images. These features are then injected into the Denoise U-Net through cross-attention mechanisms to maintain appearance consistency between the generated images and the reference image. Additionally, we utilize a pretrained Wav2Vec model as the Audio Encoder $E_A$ to extract audio embeddings, and then inject as the audio condition $c_a$ via Audio Cross Attention blocks in the Denoising U-Net, for the synchronization between motion and audio. Besides, we also integrate a Pose Encoder $E_p$ to extract keypoint maps. These keypoint maps are then concatenated with the noise latents and fed into the Denoise U-Net, serving as pose condition. Finally, we inject Temporal-Attention blocks into Denoise U-Net to capture inter-frame motion dependencies,  for motion smoothness of the animation.
\subsection{Audio-Pose Dynamic Harmonization}
In this section we introduce the core training strategy of EchoMimicV2, Audio-Pose Dynamic Harmonization(APDH). APDH is exploit to progressively reduce the condition complexity and modulate the audio and pose condition in a waltz step-like manner. This strategy consists of two main components: Pose Sampling(PS) and Audio Diffusion(AD).
\subsubsection{Pose Sampling}
\textbf{Initial Pose Phase.} We start the training of our framework with a complete pose-driven stage. Let $\mathcal{P}^{init}$ represent the complete key points maps of the half-body human figures, encoded by a Pose Encoder $\mathbf{E_{P}}$. During this stage, we mute the Audio Cross Attention blocks and only optimize other modules in the Denoising U-Net to provide a comprehensive recognition of the human actions.\\
\textbf{Iterative Pose Sampling Phase.} At the iteration level, we progressively sample some iteration steps for pose condition dropout with a iteration-increasing probability. This gradual dropout helps to mitigate over-reliance on the pose condition.\\
\textbf{Spatial Pose Sampling Phase.} At the spatial level, we sample pose condition by removing key points as the following order: the lips part first, the head second, and the body part finally. We denote the sampled pose conditions as $\mathcal{P}^{-lips}$, $\mathcal{P}^{-head}$, and $\mathcal{P}^{hands}$, respectively. By doing so, the control of the pose condition over lip movements,facial expressions, and body(respiratory rhythms) are diminished in a step-by-step manner,thereby creating space for audio-driven procedure, allowing the audio condition to play a dominant role. 

\begin{figure*}[t!]
\begin{center}
\includegraphics[width=.95\linewidth]{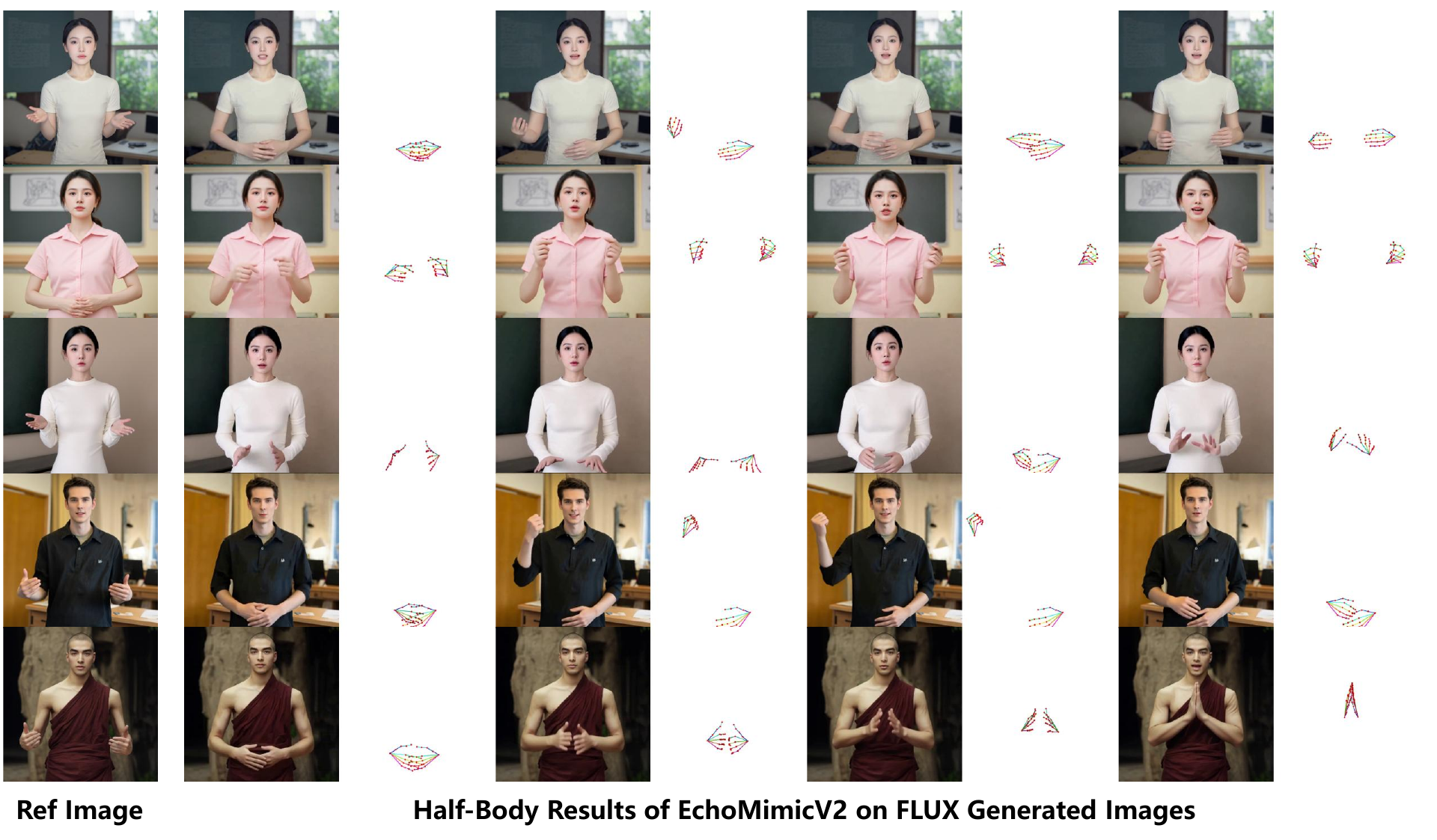}
\end{center}
\vspace{-1.5em}
   \caption{The results of EchoMimicV2 given different reference images, hand pose and audios.
}
\vspace{-1em}
\label{fig_03}
\end{figure*}
\subsubsection{Audio Diffusion}  
In the Initial Pose and Interactive Level Pose Sampling phase, the Audio Cross-Attention blocks are completely frozen. In the Spatial Level Pose Sampling phase, audio condition begins to be integrated progressively.\\
\textbf{Audio-Lips Synchronization}. After removing keypoints of lips, spatial mask $\mathcal{A}_{lips}$ for lips is applied on Audio Cross-Attention blocks as Lips Partial Attention, 
forcing audio condition solely controlling the lips movements.
By doing so, audio-lips synchronization is enhanced.\\
\textbf{Audio-Face Synchronization}. After removing keypoints of head, the
spatial mask $\mathcal{A}_{lips}$ is diffused to a head part as $\mathcal{A}_{head}$, and is applied to Audio Cross-Attention blocks as Head Partial Attention. This allows the audio condition to dominate the overall facial expression, enhancing audio-face expression synchronization.\\
\textbf{Audio-Body Correlation}. With the hands-only keypoints $\mathcal{P}^{hands}$, partial mask is diffused to the global space. This results in an entirely audio-driven half-body animation, accompanied by a pose emphasis on the hands. Note that the hands part serve as intersection between audio and pose condition.Thus, more audio cues are translated into appropriate gestures, allowing for a better capture of audio-gesture correlation.
\subsection{Head Partial Attention for Data Augmentation}
During Audio-Face Synchronization, we introduce headshot data to offset the scarcity of half-body data and enhance facial expression. To achieve this, we pad the headshot data to strictly align the spatial dimensions and head location of half-body images.
And then we reuse the Head Partial Attention to exclude the padded parts as shown in Figure \ref{fig_02}. Notably, no additional cross-attention blocks are required. To mitigate the impact of the padding for source distribution, we incorporate this step before introducing half-body data in the Audio-Face Synchronization phase.
\subsection{Temporal Modules Optimization}
Following the EchoMimic\cite{chen2024echomimic}, we integrate Temporal Cross Attention blocks into the Denoising U-Net and initialize its weights using EchoMimic. During the optimization of the Temporal modules, we use 24-frame video clips as input and freeze the other modules to ensure stability.
\subsection{Phase-specific Denoising Loss}
Besides the aforementioned condition design and training strategy, we also improve performance via the Phase-specific Denoising Loss (PhD Loss) $ L_{\text{PhD}} $.
We divide the multi-timestep denoising process into three phases based on their primary tasks: 
1) Pose-dominant phase (early) $ \mathcal{S}_{1} $; 
2) Detail-dominant phase (middle) $ \mathcal{S}_{2} $; 
3) Quality-dominant phase (final) $ \mathcal{S}_3 $. 
Accordingly, $ L_{\text{PhD}} $ includes three tailored losses, denoted as Pose-dominant Loss $ L_{\text{pose}} $, Detail-dominant Loss $ L_{\text{detail}} $, and Low-level Loss $ L_{\text{low}} $, which are applied sequentially across these three phases.\\
\textbf{Pose-dominant Loss $L_{pose}$}: Specifically, we first adopt one-step sampling to calculate the prediction $Z^{t}_0$ for the latent output $Z_{t}$ of timestep $t$ from Denoising U-Net. And then, we decode the latent $Z^{pred}_0$ by VAE decoder $D$, to obtain the RGB form $I_0^{t}$ for $Z_{t}$. Next, we employ the Pose Encoder $E_P$ to extract the keypoints map of $I_0^{t}$ and the target image $I_{target}$ as $\mathcal{M}_0^t$ and $\mathcal{M}_{target}$, respectively. Finally, we implement MSE loss on the $\mathcal{M}_p^t$ and $\mathcal{M}_p^{target}$ to calculate the $L_{pose}$ in $\mathcal{S}_1$ as follow:
\begin{equation}
\begin{aligned}
L_{pose} = MSE(\mathcal{M}_p^t, \mathcal{M}_p^{target})
\end{aligned}
\end{equation}
\vspace{-1.5em}
\\
\textbf{Detail-dominant Loss $L_{detail}$}: We use a Canny operator as used in ControlNet\cite{zhang2023adding} to extract the edges and other high-frequency details of $I_0^{t}$ and $I_{target}$ as $\mathcal{M}_d^t$ and $\mathcal{M}_d^{target}$, respectively. And then we calculate the MSE loss of $\mathcal{M}_d^t$ and $\mathcal{M}_d^{target}$ as the $L_{detail}$ in $\mathcal{S}_2$:
\begin{equation}
\begin{aligned}
L_{detail} = MSE(\mathcal{M}_d^t, \mathcal{M}_d^{target})
\end{aligned}
\end{equation}
\vspace{-.8em}
\\
\textbf{Low-level Loss $L_{low}$}: We exploit LPIPS following the Spatial Loss of EchoMimic as our $L_{low}$ in $\mathcal{S}_3$: 
\begin{equation}
\begin{aligned}
L_{detail} = LPIPS(I_0^{t}, I_{target})
\end{aligned}
\end{equation}
\vspace{-1.5em}
\\
We also follow the latent loss in LDM to optimize the model, denoted as $L_{latent}$. 
An then we obtain the overall PhD Loss as follows:\\
\begin{equation}
\begin{aligned}
L_{PhD} = \left\{ 
    \begin{aligned}
    & \lambda_{pose}\cdot L_{pose} + L_{latent} & & t \in \mathcal{S}_1 \cr 
    &\lambda_{detail}\cdot L_{detail} + L_{latent} & & t \in \mathcal{S}_2 \cr 
    &\lambda_{low}\cdot L_{low} + L_{latent} & & t \in \mathcal{S}_3
    \end{aligned}
\right.
\end{aligned}
\end{equation}
where $\lambda_{pose}$, $\lambda_{detail}$, and $\lambda_{low}$ are the loss weights.

\begin{figure}[t]
\begin{center}
\includegraphics[width=1\linewidth]{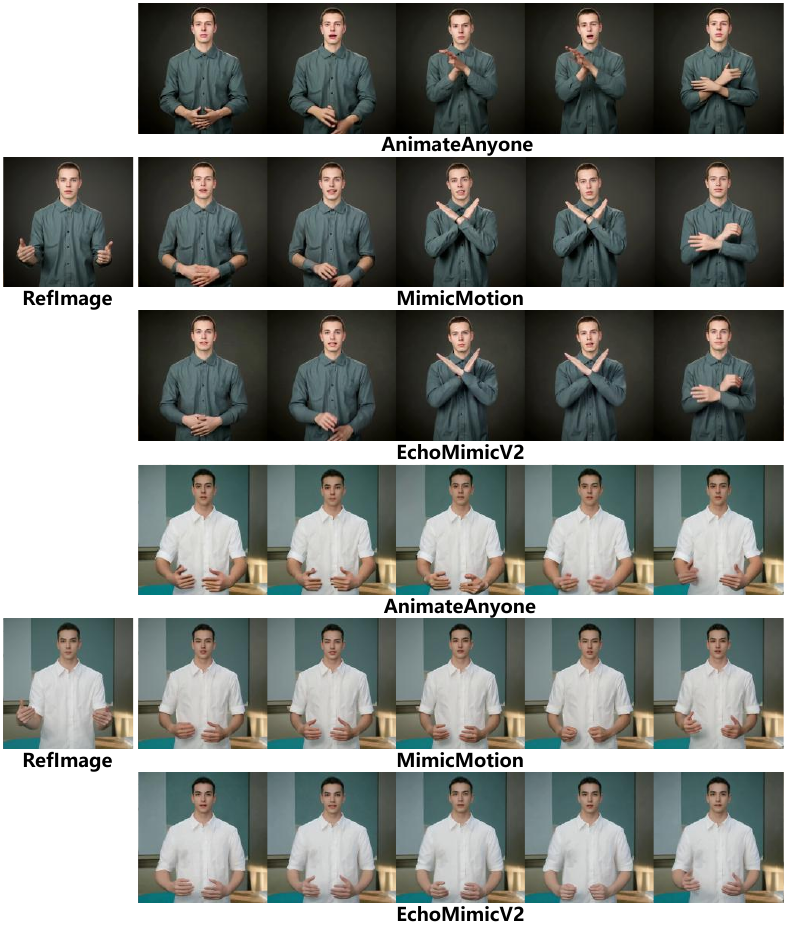}
\end{center}
\vspace{-1.5em}
   \caption{The results of EchoMimicV2 compared to pose-driven half-body human animation baselines.
}
\vspace{-1em}
\label{fig_04}
\end{figure}
\section{Experiments}
In this section, we first provide the implementation details, training datasets, evaluation benchmarks used in our experiments. Following this, we evaluate the superior performance of our approach via quantitative and qualitative comparisons with comparable methods. Next, we also perform ablation studies to analyze the effectiveness of each component of our method. 
\subsection{Experimental Setups}
\textbf{Implementation Details.} Our training procedure includes two stages. In the first stage, the model is trained to generate target frames from reference images, ensuring visual consistency.
\begin{figure}[t]
\begin{center}
\includegraphics[width=1\linewidth]{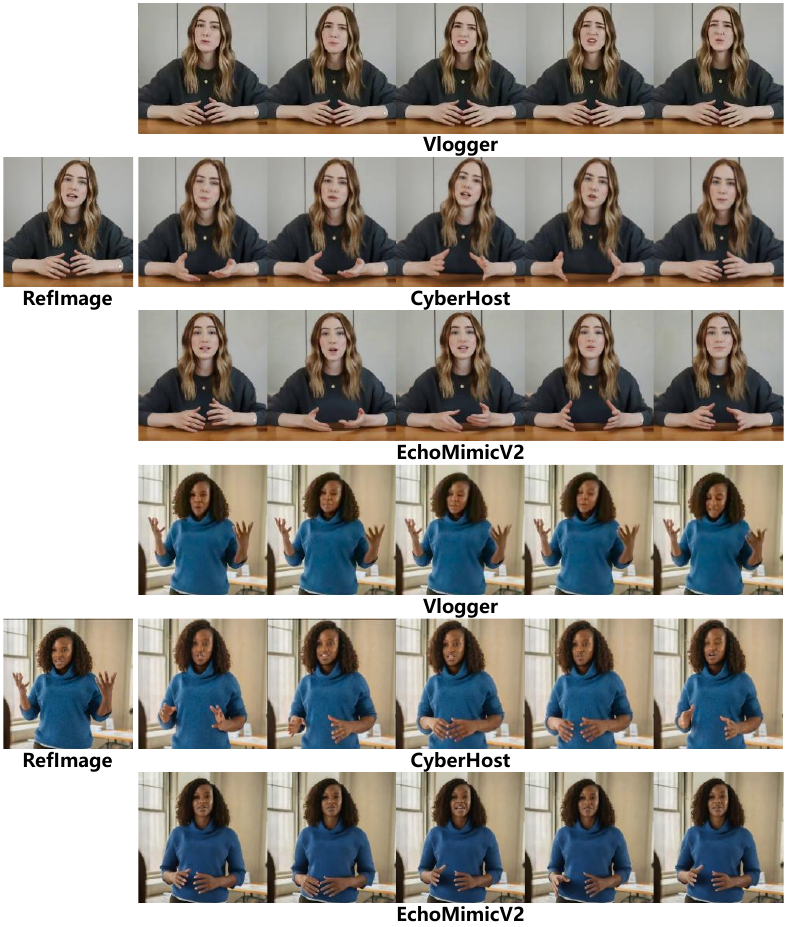}
\end{center}
\vspace{-1.5em}
   \caption{The results of EchoMimicV2 compared to audio-driven half-body human animation baselines.
}
\vspace{-1em}
\label{fig_05}
\end{figure}
We first employ complete key points maps $\mathcal{P}^{init}$ as pose condition to train the Pose Encoder and Denoising U-Net jointly, while Audio Cross Attention is masked. This process is conducted on 8 A100 GPUs for 10,000 iterations, with a batch size of 4 and a resolution of $768 \times 768$ per GPU. And then we train our framework with Audio-Pose Dynamic Harmonization. We train our framework for 10,000 steps in the Iterative Pose Sampling Phase, with the dropout probability for pose condition increasing from 0\% to 20\%. In the Spatial Pose Sampling Phase, we first train our framework by 10,000 steps utilizing $\mathcal{P}^{-lips}$ and Lips Partial Attention, and then train our framework by 10,000 steps with $\mathcal{P}^{-head}$ and Head Partial Attention on headshot data from EchoMimic. Subsequently, we train our framework by 10,000 iterations using half-body data. Finally, we conduct 10,000 iterations with $\mathcal{P}^{hands}$ and global Audio Cross Attention. For the $L_{PhD}$, the scope of $\mathcal{S}_1$ is the early 10\% timesteps, the scope of $\mathcal{S}_2$ is the following 60\% timesteps, and the rest timesteps is divided into $\mathcal{S}_3$. And the loss weights of $\lambda_{pose}$, $\lambda_{detail}$, and $\lambda_{low}$ is all set to 0.1.
In the second stage, we solely optimize the Temporal-Attention modules and other modules frozen with a batch size of 4 for 30,000 steps. For both stages, the learning rate is set to $1e^{-5}$, and the classifier-free guidance (CFG) scale for reference images and audio is set to 2.5. 
\\
\textbf{Training Datasets.}
\begin{table*}[t]
\begin{center}
\resizebox{.97\textwidth}{!}{
\begin{tabular}{l|cccc|ccc|cc|c}
\hline
Methods       & FID$\downarrow$ & FVD$\downarrow$ & SSIM$\uparrow$ & PSNR$\uparrow$ & E-FID$\downarrow$ & Sync-D$\downarrow$ & Sync-C$\uparrow$ & HKC$\uparrow$ & HKV$\uparrow$ &CSIM$\uparrow$ \\
\hline\hline
AnimateAnyone\cite{hu2024animate} &  58.98 &   1016.47  &  0.729    &  20.579 & 3.784 & 13.887 &   0.987  &   0.809  & 23.87 &  0.387 \\
MimicMotion\cite{zhang2024mimicmotion}& 53.47 &	622.62	& 0.702 & 19.278 &	2.628  &  7.958  &     1.495   &  0.907   &  24.82 &  0.526 \\
EchoMimicV2   &  \textbf{49.33}   & \textbf{598.45}  & \textbf{0.738} & \textbf{21.986} & \textcolor{blue}{2.218} & \textcolor{blue}{7.021}    &   \textcolor{blue}{7.219}  & \textcolor{red}{0.923}  & \textcolor{red}{25.28}  & \textcolor{gray}{\textbf{0.558}} \\
\hline\hline
w/o  Initial Pose &  49.99  &  602.08   &    0.730  &   21.708   &   2.235  & 6.976 & 7.019   
& 0.873  & 23.97  & 0.550 \\
w/o  Iterative Pose Sampl. &  50.01   & 605.29    &   0.727   &   21.276  & 2.276   &  6.987 & 7.005
&0.917  & 25.24  & 0.527 \\
w/o Spatial Pose Sampl. &  49.39   & 593.98    &   0.740   &   21.994   & 2.208 & 7.023  &     7.220   &   0.922  & 25.25  & 0.532 \\
\hline\hline
w/o  Headshot Data Aug. &  51.29   & 610.87    &   0.711   &   20.965   &  2.961  & 6.792 &   6.394 & 0.921  & 25.27  & 0.527 \\
w/o Audio-Lips Sync. & 49.37  & 599.82    &    0.722  &  21.988    &     2.629   &  6.803   & 6.463&  0.919  & 25.24 & 0.512 \\
w/o Audio-Face Sync.   &   51.11  &  620.75   &   0.719   &   21.837   &   2.983     &6.807 &6.286&   0.922  & 25.26  & 0.518 \\
w/o Audio-Body Corr. &  50.36   &   603.84  &    0.733  &   21.980   &   2.217     &7.025    &7.226&   0.906  & 24.98  & 0.541 \\
\hline\hline
w/o $L_{pose}$ & 51.30  & 620.27    &    0.695  &  20.783   & 2.766  & 6.954   &     7.063   & 
0.874 & 22.83  & 0.549 \\
w/o $L_{detail}$& 51.61  &  623.56  &   0.689   &   20.038  &  2.904 & 6.807  &     6.985   &   
0.896  & 24.37  & 0.541 \\
w/o $L_{low}$ &  50.86   &  602.31  &   0.675  &   19.786   &   2.226     & 7.028    &  7.223   &   0.913  & 25.20  & 0.547 \\
\hline\hline
Straightforward Baseline &  51.53   &  624.98  &   0.664  &   19.269   &   2.779     & 7.208    &  6.706   &   0.825  & 22.57  & 0.508 \\
\hline
\end{tabular}
}
\end{center}
\vspace{-1.5em}
\caption{Quantitative comparison and ablation  study of our proposed EchoMimicV2 and other SOTA methods.}
\vspace{-1.1em}
\label{table_01}
\end{table*}
Our training dataset comprises three parts: fully pose-driven dataset, half-body dataset, and headshot dataset. We use HumanVid\cite{wang2024humanvid} as 
the fully pose-driven dataset, which includes about 20,000 high-quality human-centric videos with 2D pose annotations. The half-body training dataset is curated from internet videos and focuses on half-body speaking scenarios. This dataset spans 160 hours and includes over 10,000 identities. The headshot dataset is derived from the training dataset used in EchoMimic including 540 hours of talking head videos.
\\
\textbf{Novel Half-Body Evaluation Benchmark.}
Public datasets typically evaluate audio-driven talking head or pose-driven human animation, but none specifically target audio-driven half-body animation. To fill this gap, we introduce the EchoMimicV2 Testing Dataset (EMTD) for evaluating half-body human animation. EMTD includes 65 high-definition TED videos (1080P) from YouTube, featuring 110 annotated, clear, and front-facing half-body speech segments. The open-source project will provide scripts for downloading, slicing, and processing these videos for evaluation. 
Due to limited authorization, we
only provide qualitative results and won't display visualization results. We also release characters for reference images, generated by the FLUX.1-dev \footnote{https://github.com/black-forest-labs/flux}. These resources support comparative experiments and facilitate further research and application.\\
\textbf{Evaluation Metric.} 
Half-body human animation methods are evaluated using metrics such as FID, FVD\cite{unterthiner2018towards}, PSNR\cite{hore2010image}, SSIM\cite{wang2004image}, E-FID\cite{deng2019accurate}. FID, FVD, SSIM and PSNR is the metrics to evaluate the low-level visual quality. E-FID assesses image authenticity using Inception network features, extracting expression parameters with a face reconstruction method and calculating their FID scores to gauge expression disparities. To evaluate the identity consistency, we calculate the cosine similarity
(CSIM) between the facial features of the reference image and the generated video frames. Besides, we also use the SyncNet\cite{prajwal2020lip} 
to calculate Sync-C and Sync-D for valid the audio-lip synchronization accuracy. Additionally, to evaluate the hands parts animation, HKC (Hand Keypoint Confidence) averages is adopted to evaluate hand quality in audio-driven scenarios, and HKV (Hand
Keypoint Variance) is calculated to indicate the richness of hand movements.
\subsection{Qualitative Results}
In order to assess the qualitative results of our proposed EchoMimicV2 , we use the FLUX.1-dev generate reference images, and then conduct half-body animation based on these reference images. Our approach generates high resolution half-body human videos using an audio signal, a reference image and a hand pose sequence. Figure \ref{fig_03} showcases the adaptability and resilience of our method in synthesizing a wide range of audio visual outputs with seamless synchronization to the accompanying audio. These results affirm its potential for advancing the state-of-the-art in audio-driven half-body human video generation, and illustrate
that our proposed EchoMimicV2 can be well generalized across diverse characters and intricate gestures.\\
\textbf{Comparisons with Pose-Driven Body Methods}.
We conduct the qualitative evaluation of EchoMimicV2 and compare with state-of-the-art pose-driven methods, including AnimateAnyone\cite{hu2024animate} and MimicMotion\cite{zhang2024mimicmotion}, as shown in Figure \ref{fig_05}. We can see that our proposed EchoMimicV2 outperforms the current state-of-the-art results in terms of structural integrity and identity consistency, particularly in local regions such as the hands and face. Additional video comparison results can be available in supplement material.\\
\textbf{Comparisons with Audio-Driven Body Methods}. 
Only few works, such as Vlogger\cite{zhuang2024vlogger} and CyberHost\cite{lin2024cyberhost}, support audio-driven half-body human animation. Regrettably, these methods have not been open-sourced, posing challenges for direct comparisons. We obtained relevant experimental results from the homepages of these two projects and conducted experiments using the same reference image. As depicted in Figure \ref{fig_04}, our proposed EchoMimicV2 surpasses Vlogger and CyberHost in both generated image quality and the naturalness of movements.
\vspace{-0.5em}
\subsection{Quantitative Results}
As illustrated in Table \ref{table_01}, we quantitatively compare EchoMimicV2 with state-of-the-art human animation methods, including AnimateAnyone\cite{hu2024animate} and MimicMotion\cite{zhang2024mimicmotion}. The results demonstrate that EchoMimicV2 significantly improves overall performance. Specifically, EchoMimicV2 shows substantial enhancements in quality metrics (FID, FVD, SSIM, and PSNR) and competitive results in synchronization metrics (Sync-C and Sync-D). Additionally, EchoMimicV2 surpasses other SOTA methods in the consistency metric (CSIM). Note that EchoMimicV2 achieves a new SOTA in hand related quality metrics (HKV, HKC).
\subsection{Ablation Study}
We analyze the effectiveness of our Audio-Pose Dynamic Harmonization (APDH) strategy, condition simplifying design and PhD Loss in Table \ref{table_01}.\\
\textbf{Analysis for Pose Sampling.}
We conduct ablation studies to validate the effectiveness of each phase of Pose Sampling. Specifically, we can observe that Initial Pose and Iterative Pose Sampling phase contribute to overall performance improvement, due to the strong motion prior provided by the full-body pose condition. On the other hand, the row $7$ of Table \ref{table_01} shows that complete pose condition (without Spatial Pose Sampling) has limited impact on each metric, indicating the redundancy of full-body pose, and our APDH strategy can still achieve stable animation with hands-only pose condition.\\
\textbf{Analysis for Audio Diffusion.}
As shown in Table \ref{table_01}, the rows $9$ and $10$ indicate that Audio Diffusion enhances lip movements and facial expressiveness. Furthermore, row $10$ demonstrates that the audio condition also improves body and hand animation, indicating that EchoMimicV2 captures the fine-grained correlation, that is, audio-related breathing rhythm and audio-related gesture.\\
\textbf{Analysis for Headshot Data Augmentation.}
The row $8$ in Table \ref{table_01} shows that headshot data augmentation significantly impacts synchronization metrics (Sync-D, Sync-C).\\
\textbf{Analysis for PhD Loss.}
We also validate the impact of each component in the PhD Loss for EchoMimicV2. As shown in rows $12$ to $14$ of Table \ref{table_01}, we observe that $L_{pose}$ has a significant impact on overall metrics because it compensates for the incompleteness of pose condition. Additionally, $L_{detail}$ significantly boosts performance on local quality metrics such as Sync-C, Sync-D, E-FID, HKC, and HKV. Moreover, $L_{low}$ contributes to performance improvement in quality metrics (SSIM, PSNR).\\
\textbf{Analysis for Straightforward Baseline.}
We also conducted an ablation study to validate the effectiveness of the overall APDH training strategy and PhD Loss. We trained the backbone with half-body audio and hand pose conditions, without the progressive APDH strategy and PhD Loss, as a straightforward baseline. As shown in the last row of Table \ref{table_01}, this baseline achieves suboptimal results due to inadequate transitions between the two conditions. These results demonstrate that the APDH and PhD Loss are important for stable training with simplified conditions. \\
\textbf{Analysis for Hands Pose Condition}. 
\begin{figure}[t]
\begin{center}
\includegraphics[width=.97\linewidth]{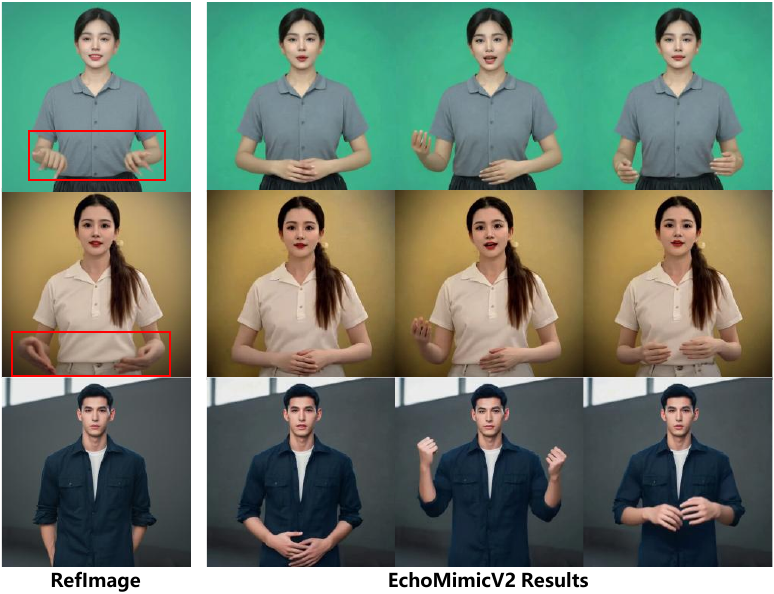}
\end{center}
\vspace{-1.5em}
   \caption{High-fidelity hands generation of EchoMimicV2 when no hands or deformed hands in RefImage.
}
\vspace{-1.5em}
\label{fig_06}
\end{figure}
Current image and video generation methods face intrinsic challenges in generating detailed hand regions. As shown in Figure \ref{fig_06}, even state-of-the-art text-to-image methods (\eg FLUX.1) struggle with hand synthesis, an issue magnified in audio-driven generation due to the weak correlation between audio signals and motion. To address this, EchoMimicV2 combines hand pose and audio conditions, demonstrating strong hand repair capabilities despite the low proportion of hand pixels in half-body images. Remarkably, EchoMimicV2 generates high-fidelity hands even when they are absent in the reference images as shown in Figure \ref{fig_06}. By removing all pose conditions and maintaining other settings, we achieve a fully audio-driven animation model that captures general gestures, correlating hand rhythm with audio tones. However, this model cannot generate specific gestures like clenching fists or saluting. Due to space constraints, video results are provided in the supplementary material.

\vspace{-.7em}
\section{Limitation}
This paper presents significant advancements in audio-driven half-body human animation, but it is important to acknowledge existing limitations and areas for further improvement.
(1) Audio to Hand Pose Generation: The proposed method  requires predefined hand pose sequences, relying on human input for high-quality animation, limiting practical applications. Future work will explore generating hand pose sequences directly from audio to in an end-to-end paradigm.
(2) Animation for Non-cropped Reference Image: While EchoMimicV2 performs robustly on cropped half-body images, its performance declines on non-cropped images, such as full-body images.
\vspace{-0.5em}
\section{Conclusion}
In this work, we propose an effective EchoMimicV2 framework to generate striking half-body human animation driven by simplified conditions. We achieve audio-pose
condition collaboration and pose condition
simplification through our proposed APDH training strategy and timesteps-specific PhD Loss, while seamlessly augmenting facial expressiveness via HPA. Comprehensive experiments demonstrate that EchoMimicV2 exceeds current state-of-the-art techniques in both quantitative and qualitative results. Furthermore, we introduce a novel benchmark for evaluating half-body human animation. For the advancement of the community, we make our source code and test dataset available for open-source use.
\section{Supplementary Material}

\noindent \textcolor{red}{\textbf{Notification.1}}. Generated video results can bee see on \texttt{echomimicv2.html} in the archive file named 
\begin{figure}[t]
\begin{center}
\includegraphics[width=1.\linewidth]{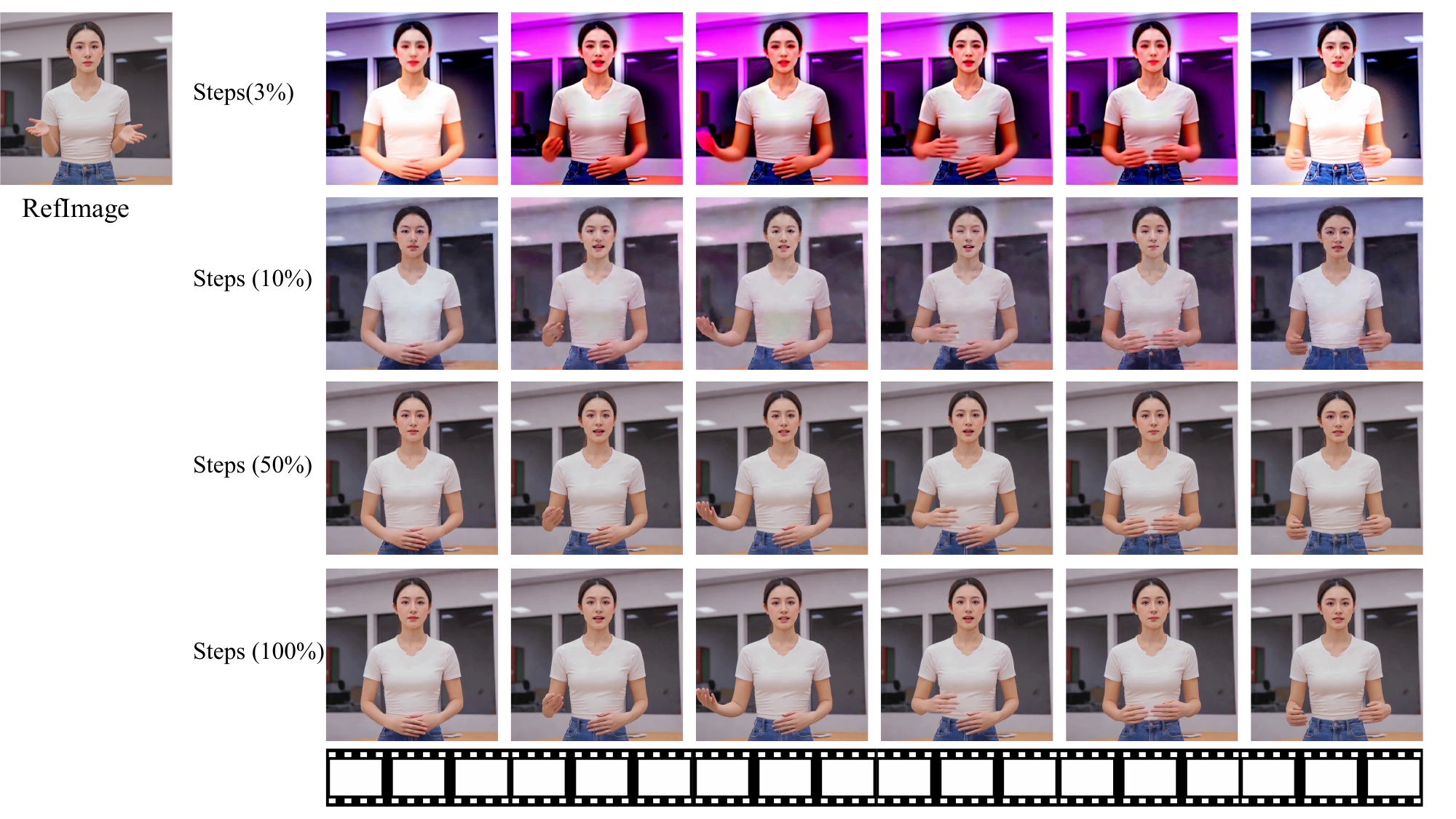}
\end{center}
\vspace{-1.5em}
   \caption{The visualization for intermediate results of multi-timesteps denoising. The motion, details and low-leval quality is timestep-by-timestep optimized in the EchoMimicV2 training.
}
\label{suppl_fig_01}
\end{figure}
\texttt{echomimic\_v2\_suppl.zip}.\\
\textcolor{red}{\textbf{Notification.2}}. We provide the file named \texttt{EMDT.txt} 
\begin{figure}[t]
\begin{center}
\includegraphics[width=1.\linewidth]{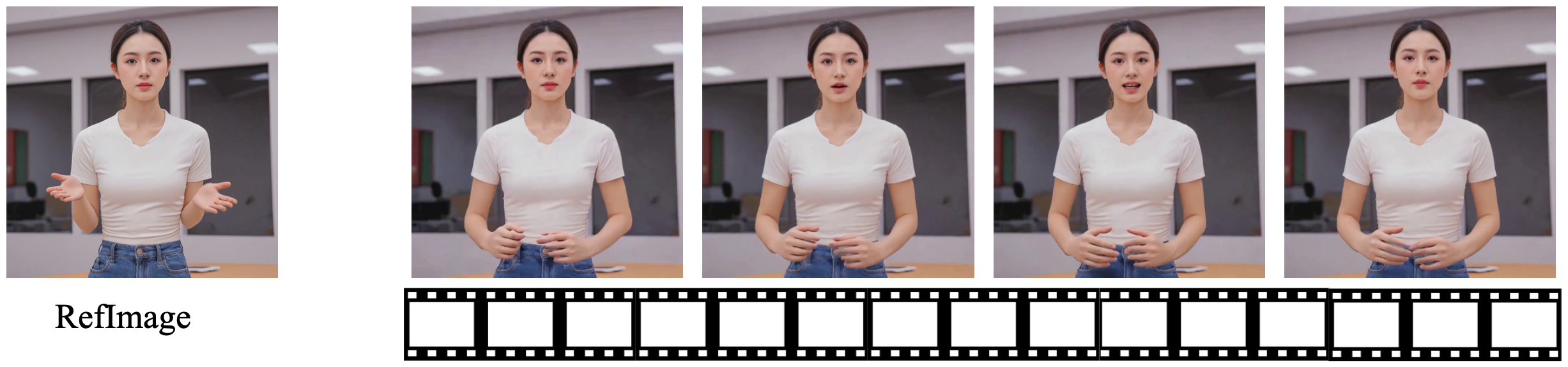}
\end{center}
\vspace{-1.5em}
   \caption{Results without hand pose condition. On the one hand,  only the rhythm of hands is learned by EchoMimicV2 (fine-tuned), but the range of motion is restricted. On the other hand, the local quality of hands is well learned.
}
\label{suppl_fig_02}
\end{figure}
in \texttt{echomimic\_v2\_suppl.zip} for URL for EMDT dataset.\\
\noindent\textbf{Analysis for PhD Loss}.
Our proposed PhD Loss is aimed to enhance motion synchronization, details and low-level quality for Pose-dominant, Detail-dominant and Quality-dominate phase, respectively. To validate the effectiveness of the phase-specific design of PhD Loss, we visualize the generation for the three denoising phase, respectively. Specifically, 1) we random select the timesteps for the three phase ($t=30,150,500,1000$); 3) And then we use one-step sampling to get one-step perdition;
3) Next, we decode the output noise latent via VAE Decoder to obtain the phase-specific generated videos as shown in Figure \ref{suppl_fig_01}. We can see that the early phase, the pose-audio synchronization has been achieved, but details such as identity, and low-level quality is not good. however, the details are learned well in the middle phase. Finally, the low-level quality is improved in the final phase. The visualization validates the motivation of our proposed PhD Loss.
\\
\textbf{Analysis for Non-Pose Driven Animation}. To validate that our EchoMimicV2 effectively learns the correlation between audio and gesture movements, and to analyze the impact of hand-only pose condition, we fine-tune the model without pose condition, and visualize the generation results as shown in Figure \ref{suppl_fig_02}. We can see that although omitting pose information, high image quality and detail for hands can still be learned, but the range of motion for hands is restricted. Hence, the hands-only pose condition can improve the diversity of gesture movements such as clenching fists or saluting (which can be see in our project page in the supplementary material).\\
\textbf{More Details of PhD Loss}. Considering the issue of back propagation, Gaussian maps $\mathcal{M}_p^t$ and $\mathcal{M}_p^{target}$ is utilized in the $L_{pose}$. While in the $L_{detail}$, the Canny operator is also modified via remove non-differentiable parts and is converted into a fixed convolutional kernel to ensure differentiability.

{
    \small
    \bibliographystyle{ieeenat_fullname}
    \bibliography{main}

@String(ICCV= {Int. Conf. Comput. Vis.})

@String(ECCV= {Eur. Conf. Comput. Vis.})

@String(AAAI = {AAAI})

@String(ICCV  = {ICCV})

@String(ECCV  = {ECCV})

@article{dhariwal2021diffusion,
  title={Diffusion models beat gans on image synthesis},
  author={Dhariwal, Prafulla and Nichol, Alexander},
  journal={Advances in neural information processing systems},
  volume={34},
  pages={8780--8794},
  year={2021}
}

@article{ho2020denoising,
  title={Denoising diffusion probabilistic models},
  author={Ho, Jonathan and Jain, Ajay and Abbeel, Pieter},
  journal={Advances in neural information processing systems},
  volume={33},
  pages={6840--6851},
  year={2020}
}

@inproceedings{rombach2022high,
  title={High-resolution image synthesis with latent diffusion models},
  author={Rombach, Robin and Blattmann, Andreas and Lorenz, Dominik and Esser, Patrick and Ommer, Bj{\"o}rn},
  booktitle={Proceedings of the IEEE/CVF conference on computer vision and pattern recognition},
  pages={10684--10695},
  year={2022}
}

@article{guo2023animatediff,
  title={Animatediff: Animate your personalized text-to-image diffusion models without specific tuning},
  author={Guo, Yuwei and Yang, Ceyuan and Rao, Anyi and Liang, Zhengyang and Wang, Yaohui and Qiao, Yu and Agrawala, Maneesh and Lin, Dahua and Dai, Bo},
  journal={arXiv preprint arXiv:2307.04725},
  year={2023}
}

@article{chen2023videocrafter1,
  title={Videocrafter1: Open diffusion models for high-quality video generation},
  author={Chen, Haoxin and Xia, Menghan and He, Yingqing and Zhang, Yong and Cun, Xiaodong and Yang, Shaoshu and Xing, Jinbo and Liu, Yaofang and Chen, Qifeng and Wang, Xintao and others},
  journal={arXiv preprint arXiv:2310.19512},
  year={2023}
}

@article{blattmann2023stable,
  title={Stable video diffusion: Scaling latent video diffusion models to large datasets},
  author={Blattmann, Andreas and Dockhorn, Tim and Kulal, Sumith and Mendelevitch, Daniel and Kilian, Maciej and Lorenz, Dominik and Levi, Yam and English, Zion and Voleti, Vikram and Letts, Adam and others},
  journal={arXiv preprint arXiv:2311.15127},
  year={2023}
}

@inproceedings{esser2023structure,
  title={Structure and content-guided video synthesis with diffusion models},
  author={Esser, Patrick and Chiu, Johnathan and Atighehchian, Parmida and Granskog, Jonathan and Germanidis, Anastasis},
  booktitle={Proceedings of the IEEE/CVF International Conference on Computer Vision},
  pages={7346--7356},
  year={2023}
}

@inproceedings{yang2023rerender,
  title={Rerender a video: Zero-shot text-guided video-to-video translation},
  author={Yang, Shuai and Zhou, Yifan and Liu, Ziwei and Loy, Chen Change},
  booktitle={SIGGRAPH Asia 2023 Conference Papers},
  pages={1--11},
  year={2023}
}

@inproceedings{hu2024animate,
  title={Animate anyone: Consistent and controllable image-to-video synthesis for character animation},
  author={Hu, Li},
  booktitle={Proceedings of the IEEE/CVF Conference on Computer Vision and Pattern Recognition},
  pages={8153--8163},
  year={2024}
}

@article{zhang2024mimicmotion,
  title={Mimicmotion: High-quality human motion video generation with confidence-aware pose guidance},
  author={Zhang, Yuang and Gu, Jiaxi and Wang, Li-Wen and Wang, Han and Cheng, Junqi and Zhu, Yuefeng and Zou, Fangyuan},
  journal={arXiv preprint arXiv:2406.19680},
  year={2024}
}

@misc{brooks2024video,
  title={Video generation models as world simulators},
  author={Brooks, Tim and Peebles, Bill and Holmes, Connor and DePue, Will and Guo, Yufei and Jing, Li and Schnurr, David and Taylor, Joe and Luhman, Troy and Luhman, Eric and others},
  journal={2024-03-03]. https://openai. com/research/video-generation-modelsas-world-simulators},
  year={2024}
}

@inproceedings{chang2023magicpose,
  title={MagicPose: Realistic Human Poses and Facial Expressions Retargeting with Identity-aware Diffusion},
  author={Chang, Di and Shi, Yichun and Gao, Quankai and Xu, Hongyi and Fu, Jessica and Song, Guoxian and Yan, Qing and Zhu, Yizhe and Yang, Xiao and Soleymani, Mohammad},
  booktitle={Forty-first International Conference on Machine Learning},
  year={2023}
}

@article{tu2024motionfollower,
  title={MotionFollower: Editing Video Motion via Lightweight Score-Guided Diffusion},
  author={Tu, Shuyuan and Dai, Qi and Zhang, Zihao and Xie, Sicheng and Cheng, Zhi-Qi and Luo, Chong and Han, Xintong and Wu, Zuxuan and Jiang, Yu-Gang},
  journal={arXiv preprint arXiv:2405.20325},
  year={2024}
}

@article{wang2024unianimate,
  title={UniAnimate: Taming Unified Video Diffusion Models for Consistent Human Image Animation},
  author={Wang, Xiang and Zhang, Shiwei and Gao, Changxin and Wang, Jiayu and Zhou, Xiaoqiang and Zhang, Yingya and Yan, Luxin and Sang, Nong},
  journal={arXiv preprint arXiv:2406.01188},
  year={2024}
}

@inproceedings{yang2023effective,
  title={Effective whole-body pose estimation with two-stages distillation},
  author={Yang, Zhendong and Zeng, Ailing and Yuan, Chun and Li, Yu},
  booktitle={Proceedings of the IEEE/CVF International Conference on Computer Vision},
  pages={4210--4220},
  year={2023}
}

@article{wang2024v,
  title={V-Express: Conditional Dropout for Progressive Training of Portrait Video Generation},
  author={Wang, Cong and Tian, Kuan and Zhang, Jun and Guan, Yonghang and Luo, Feng and Shen, Fei and Jiang, Zhiwei and Gu, Qing and Han, Xiao and Yang, Wei},
  journal={arXiv preprint arXiv:2406.02511},
  year={2024}
}

@article{wei2024aniportrait,
  title={Aniportrait: Audio-driven synthesis of photorealistic portrait animation},
  author={Wei, Huawei and Yang, Zejun and Wang, Zhisheng},
  journal={arXiv preprint arXiv:2403.17694},
  year={2024}
}

@article{xu2024hallo,
  title={Hallo: Hierarchical audio-driven visual synthesis for portrait image animation},
  author={Xu, Mingwang and Li, Hui and Su, Qingkun and Shang, Hanlin and Zhang, Liwei and Liu, Ce and Wang, Jingdong and Yao, Yao and Zhu, Siyu},
  journal={arXiv preprint arXiv:2406.08801},
  year={2024}
}

@inproceedings{zhang2023sadtalker,
  title={Sadtalker: Learning realistic 3d motion coefficients for stylized audio-driven single image talking face animation},
  author={Zhang, Wenxuan and Cun, Xiaodong and Wang, Xuan and Zhang, Yong and Shen, Xi and Guo, Yu and Shan, Ying and Wang, Fei},
  booktitle={Proceedings of the IEEE/CVF Conference on Computer Vision and Pattern Recognition},
  pages={8652--8661},
  year={2023}
}

@article{lin2024cyberhost,
  title={CyberHost: Taming Audio-driven Avatar Diffusion Model with Region Codebook Attention},
  author={Lin, Gaojie and Jiang, Jianwen and Liang, Chao and Zhong, Tianyun and Yang, Jiaqi and Zheng, Yanbo},
  journal={arXiv preprint arXiv:2409.01876},
  year={2024}
}

@article{chen2024echomimic,
  title={Echomimic: Lifelike audio-driven portrait animations through editable landmark conditions},
  author={Chen, Zhiyuan and Cao, Jiajiong and Chen, Zhiquan and Li, Yuming and Ma, Chenguang},
  journal={arXiv preprint arXiv:2407.08136},
  year={2024}
}

@inproceedings{zhang2023adding,
  title={Adding conditional control to text-to-image diffusion models},
  author={Zhang, Lvmin and Rao, Anyi and Agrawala, Maneesh},
  booktitle={Proceedings of the IEEE/CVF International Conference on Computer Vision},
  pages={3836--3847},
  year={2023}
}

@inproceedings{qiao2017real,
  title={Real-time human gesture grading based on OpenPose},
  author={Qiao, Sen and Wang, Yilin and Li, Jian},
  booktitle={2017 10th International Congress on Image and Signal Processing, BioMedical Engineering and Informatics (CISP-BMEI)},
  pages={1--6},
  year={2017},
  organization={IEEE}
}

@inproceedings{karras2023dreampose,
  title={Dreampose: Fashion image-to-video synthesis via stable diffusion},
  author={Karras, Johanna and Holynski, Aleksander and Wang, Ting-Chun and Kemelmacher-Shlizerman, Ira},
  booktitle={2023 IEEE/CVF International Conference on Computer Vision (ICCV)},
  pages={22623--22633},
  year={2023},
  organization={IEEE}
}

@inproceedings{xu2024magicanimate,
  title={Magicanimate: Temporally consistent human image animation using diffusion model},
  author={Xu, Zhongcong and Zhang, Jianfeng and Liew, Jun Hao and Yan, Hanshu and Liu, Jia-Wei and Zhang, Chenxu and Feng, Jiashi and Shou, Mike Zheng},
  booktitle={Proceedings of the IEEE/CVF Conference on Computer Vision and Pattern Recognition},
  pages={1481--1490},
  year={2024}
}

@inproceedings{guler2018densepose,
  title={Densepose: Dense human pose estimation in the wild},
  author={G{\"u}ler, R{\i}za Alp and Neverova, Natalia and Kokkinos, Iasonas},
  booktitle={Proceedings of the IEEE conference on computer vision and pattern recognition},
  pages={7297--7306},
  year={2018}
}

@article{shao2024human4dit,
  title={Human4DiT: Free-view Human Video Generation with 4D Diffusion Transformer},
  author={Shao, Ruizhi and Pang, Youxin and Zheng, Zerong and Sun, Jingxiang and Liu, Yebin},
  journal={arXiv preprint arXiv:2405.17405},
  year={2024}
}

@inproceedings{bogo2016keep,
  title={Keep it SMPL: Automatic estimation of 3D human pose and shape from a single image},
  author={Bogo, Federica and Kanazawa, Angjoo and Lassner, Christoph and Gehler, Peter and Romero, Javier and Black, Michael J},
  booktitle={Computer Vision--ECCV 2016: 14th European Conference, Amsterdam, The Netherlands, October 11-14, 2016, Proceedings, Part V 14},
  pages={561--578},
  year={2016},
  organization={Springer}
}

@inproceedings{zhuang2024vlogger,
  title={Vlogger: Make your dream a vlog},
  author={Zhuang, Shaobin and Li, Kunchang and Chen, Xinyuan and Wang, Yaohui and Liu, Ziwei and Qiao, Yu and Wang, Yali},
  booktitle={Proceedings of the IEEE/CVF Conference on Computer Vision and Pattern Recognition},
  pages={8806--8817},
  year={2024}
}

@article{yang2024megactor,
  title={MegActor-$\Sigma$: Unlocking Flexible Mixed-Modal Control in Portrait Animation with Diffusion Transformer},
  author={Yang, Shurong and Li, Huadong and Wu, Juhao and Jing, Minhao and Li, Linze and Ji, Renhe and Liang, Jiajun and Fan, Haoqiang and Wang, Jin},
  journal={arXiv preprint arXiv:2408.14975},
  year={2024}
}

@article{tian2024emo,
  title={Emo: Emote portrait alive-generating expressive portrait videos with audio2video diffusion model under weak conditions},
  author={Tian, Linrui and Wang, Qi and Zhang, Bang and Bo, Liefeng},
  journal={arXiv preprint arXiv:2402.17485},
  year={2024}
}

@article{liu2024tango,
  title={TANGO: Co-Speech Gesture Video Reenactment with Hierarchical Audio Motion Embedding and Diffusion Interpolation},
  author={Liu, Haiyang and Yang, Xingchao and Akiyama, Tomoya and Huang, Yuantian and Li, Qiaoge and Kuriyama, Shigeru and Taketomi, Takafumi},
  journal={arXiv preprint arXiv:2410.04221},
  year={2024}
}

@article{kingma2013auto,
  title={Auto-encoding variational bayes},
  author={Kingma, Diederik P},
  journal={arXiv preprint arXiv:1312.6114},
  year={2013}
}

@article{wang2024humanvid,
  title={HumanVid: Demystifying Training Data for Camera-controllable Human Image Animation},
  author={Wang, Zhenzhi and Li, Yixuan and Zeng, Yanhong and Fang, Youqing and Guo, Yuwei and Liu, Wenran and Tan, Jing and Chen, Kai and Xue, Tianfan and Dai, Bo and others},
  journal={arXiv preprint arXiv:2407.17438},
  year={2024}
}

@inproceedings{prajwal2020lip,
  title={A lip sync expert is all you need for speech to lip generation in the wild},
  author={Prajwal, KR and Mukhopadhyay, Rudrabha and Namboodiri, Vinay P and Jawahar, CV},
  booktitle={Proceedings of the 28th ACM international conference on multimedia},
  pages={484--492},
  year={2020}
}

@article{wang2004image,
  title={Image quality assessment: from error visibility to structural similarity},
  author={Wang, Zhou and Bovik, Alan C and Sheikh, Hamid R and Simoncelli, Eero P},
  journal={IEEE transactions on image processing},
  volume={13},
  number={4},
  pages={600--612},
  year={2004},
  publisher={IEEE}
}

@inproceedings{hore2010image,
  title={Image quality metrics: PSNR vs. SSIM},
  author={Hore, Alain and Ziou, Djemel},
  booktitle={2010 20th international conference on pattern recognition},
  pages={2366--2369},
  year={2010},
  organization={IEEE}
}

@article{unterthiner2018towards,
  title={Towards accurate generative models of video: A new metric \& challenges},
  author={Unterthiner, Thomas and Van Steenkiste, Sjoerd and Kurach, Karol and Marinier, Raphael and Michalski, Marcin and Gelly, Sylvain},
  journal={arXiv preprint arXiv:1812.01717},
  year={2018}
}

@inproceedings{deng2019accurate,
  title={Accurate 3d face reconstruction with weakly-supervised learning: From single image to image set},
  author={Deng, Yu and Yang, Jiaolong and Xu, Sicheng and Chen, Dong and Jia, Yunde and Tong, Xin},
  booktitle={Proceedings of the IEEE/CVF conference on computer vision and pattern recognition workshops},
  pages={0--0},
  year={2019}
}

@inproceedings{meng2022attention,
  title={Attention diversification for domain generalization},
  author={Meng, Rang and Li, Xianfeng and Chen, Weijie and Yang, Shicai and Song, Jie and Wang, Xinchao and Zhang, Lei and Song, Mingli and Xie, Di and Pu, Shiliang},
  booktitle={European conference on computer vision},
  pages={322--340},
  year={2022},
  organization={Springer}
}

@inproceedings{meng2022slimmable,
  title={Slimmable domain adaptation},
  author={Meng, Rang and Chen, Weijie and Yang, Shicai and Song, Jie and Lin, Luojun and Xie, Di and Pu, Shiliang and Wang, Xinchao and Song, Mingli and Zhuang, Yueting},
  booktitle={Proceedings of the IEEE/CVF Conference on Computer Vision and Pattern Recognition},
  pages={7141--7150},
  year={2022}
}

@inproceedings{meng2020neural,
  title={Neural inheritance relation guided one-shot layer assignment search},
  author={Meng, Rang and Chen, Weijie and Xie, Di and Zhang, Yuan and Pu, Shiliang},
  booktitle={Proceedings of the AAAI Conference on Artificial Intelligence},
  volume={34},
  number={04},
  pages={5158--5165},
  year={2020}
}

@inproceedings{ignatov2018pirm,
  title={Pirm challenge on perceptual image enhancement on smartphones: Report},
  author={Ignatov, Andrey and Timofte, Radu and Van Vu, Thang and Minh Luu, Tung and X Pham, Trung and Van Nguyen, Cao and Kim, Yongwoo and Choi, Jae-Seok and Kim, Munchurl and Huang, Jie and others},
  booktitle={Proceedings of the European Conference on Computer Vision (ECCV) Workshops},
  pages={0--0},
  year={2018}
}
}


\end{document}